\documentclass[smallextended]{svjour3} 

\usepackage{graphicx}
\usepackage{longtable}

\usepackage[version=3]{mhchem}
\usepackage{multirow}
\usepackage{array}
\usepackage{bm}
\usepackage{amsmath}
\usepackage{amssymb}\usepackage{lscape}
\usepackage{pdfpages}
\usepackage{mathtools, cuted}
\usepackage{gensymb}
\usepackage[T1]{fontenc}

\title{ 
Clifford Boundary Conditions for Periodic 
Systems: the Madelung Constant of
Cubic Crystals in 1, 2 and 3 Dimensions
} 

\author{Nicolas Tavernier \and Gian Luigi Bendazzoli \and V\'eronique Brumas \and Stefano Evangelisti \and J. Arjan Berger}
\institute{\lcpq \and \unibo \and \etsf}

\institute{Nicolas Tavernier  \and V\'eronique Brumas \and Stefano Evangelisti \at
              Laboratoire de Chimie et Physique Quantiques, CNRS, Universit\'e de Toulouse, UPS \\
           \and
           Gian Luigi Bendazzoli  \at
              Universit\`a di Bologna, Bologna, Italy \\
            \and
            J. Arjan Berger\at
            Laboratoire de Chimie et Physique Quantiques and uropean Theoretical Spectroscopy Facility (ETSF), IRSAMC, CNRS, Universit\'e de Toulouse, UPS  \\
            \email{arjan.berger@irsamc.ups-tlse.fr} 
}

\date{\today}
\begin{document}

\begin{abstract}
In this work we demonstrate the robustness of a real-space approach for the treatment of infinite systems 
described with periodic boundary conditions that we have recently proposed [J. Phys. Chem. Lett. 17, 7090].
In our approach we extract a fragment, i.e., a supercell, out of the infinite system, and
then modifying its topology into the that of a Clifford torus which is a flat, finite and border-less manifold.
We then renormalize the distance between two points by defining it as the Euclidean distance in the embedding space
of the Clifford torus.
With our method we have been able to calculate the reference results available 
in the literature with a remarkable accuracy, and at a very low computational effort.
In this work we show that our approach is robust with respect to the shape of the supercell.
In particular, we show that the Madelung constants converge to the same values but that the convergence properties are different.
Our approach scales linearly with the number of atoms.
The calculation of Madelung constants only takes a few seconds on a laptop computer for a relative precision of about 10$^{-6}$.
\end{abstract}

\maketitle

\renewcommand{\baselinestretch}{1.5}

\section{Introduction}

\vspace{4mm}
The formalism of periodic boundary conditions (PBC) is widely used in solid-state physics, chemistry and material sciences,  
in order to simulate very large crystalline systems.
The PBC formalism is often applied by employing Born - Von K\'arm\'an boundary conditions, originally
developed in the framework of the theory of elasticity and specific heat of solids. 
However, in this case we are in presence of very short-range interactions, 
so the global topology of the whole system is not really an important issue.
On the other hand, in order to deal with the exact treatment of the electronic 
structure of crystalline systems, the long-range nature of the Coulomb interaction completely changes the situation.
In fact, the infinite sums needed to compute the total energy of a Coulomb system are,
taken term by term, invariably conditionally convergent series, 
for which the order of summation has a crucial effect on the final result.
In particular, the Riemann series theorem holds, which says that 
if an infinite series of real numbers is conditionally convergent, then the terms of the series
can be rearranged in such a way that the new series converges to an arbitrary real number \cite{rudin_principles_1964}. 
%
%

For this reason, for more than a century, 
a large number of sophisticated techniques has been developed to deal with
this problem, whose presence occurs both in classical and quantum theory.
These approaches go back to the 
Evjen \cite{evjen_stability_1932} and Ewald \cite{ewald_berechnung_1921} techniques,
whose works gave rise to the two most common types of approaches used 
to compute Coulomb sums in the case of infinite systems, namely direct-summation and integral techniques, respectively. 
In particular, there exist many variations and generalizations of the Ewald method, e.g., the particle-mesh Ewald (PME) method~\cite{Darden_1993} and smooth PME~\cite{Essmann_1995}, the particle-particle-particle-mesh (P$^3$M) approach~\cite{Eastwood_1984}, 
the parallel three-dimensional nonequispaced fast fourier transform technique (P$^2$NFFT)~\cite{Pippig_2013}, and the spectrally accurate Ewald method~\cite{Lindbo_2012}.
Finally, we should also mention the fast multipole method (FMM)~\cite{Rokhlin_1985} for calculating Coulomb sums.
However, although these methods are usually highly effective, they are also very technical and, therefore, their implementation is not straightforward.
Moreover, the technicalities of these methods hide the physical picture of the problem.
Finally, these approaches depend on parameters and their efficiency can be sensitive to the choice of these parameters.
Therefore we have recently developed a very simple direct-sum approach without parameters that can efficiently perform lattice sums~\cite{Tavernier_2020}.

In this article, we illustrate the robustness of our approach.
In particular, we show that the infinite-size limit does not depend on the ratios between the supercell dimensions,
provided it is a square parallelepiped.
In other words, it is not needed that the edges of the cell are equal, even for cubic crystals, in order to converge to the correct value.
In this way, to a given threshold of accuracy, any crystal system can be treated within this formalism.
The proposed method originated in the more general context of defining a position operator for periodic systems
\cite{valenca_ferreira_de_aragao_simple_2019}.
It consists essentially in the extraction of a fragment, i.e., a supercell, out of the infinite system, and
then modifying its topology into the topology of a torus.
However, in order to conserve the crystal regular structure, and in particular, the angles, 
of the original system, a special type of torus is to be used: it is a Clifford torus
${\cal T}^n$ of dimension $n$:
this is a {\em flat border-less manifold}, first described by 
W.K. Clifford during a meeting of the British Association
for the Advancement of Sciences, in September 1873.
A Clifford torus which is able to contain a 2-dimensional (or, respectively, 
3-dimensional) supercell
is a {\em flat} closed surface (respectively, a volume), embedded in a 2-D (or 3-D) 
{\em complex} Euclidean space.
We note that, alternatively, one can also embed the supercell in real space at the price 
of doubling the number of spatial dimensions.

However, despite this difficulty, they are perfectly suited 
to represent a fragment of the real, infinite system.
Unlike the $n$-dimensional sphere, ${\cal S}^n$, the n-dimensional torus ${\cal T}^n$ has zero 
intrinsic curvature, and it is therefore
more suited to host a fragment of a crystal without deforming it.

Once the topology of the supercell has been redefined in such a way,
we have to introduce the notion of distance between two arbitrary points in this manifold.
We note that the Clifford torus is a Riemannian manifold, 
and therefore a Riemannian Metric is naturally defined on it.
However, such a metric is not a unique function of the two points, and in general
its derivative is not even continuous.
For these reasons, it is not suitable as the distance 
to be used in dealing with the Coulomb interaction.
But a second distance naturally arises for the torus, 
because of the embedding property of this manifold:
it is the Euclidean distance defined in the embedding space
(either real or complex),
and it is this single-valued and smooth distance that we will use to define the Coulomb potential.

At the end of the whole procedure, the supercell extracted from the original
infinite system has been transformed into
a toroidal manifold in which all the atoms that were equivalent in the original crystal are still equivalent.
We note, in particular, that the angles among atoms of the supercell have not been modified with respect
to the corresponding angles in the infinite crystal.
This is a crucial advantage of using a Clifford torus:
because of its flatness, this manifold can host a 2-D or 3-D supercell without deforming it,
as it would happen in the case of ordinary tori, or spheres.
The Euclidean distance introduced between two points in the supercell is a unique and  smooth 
function of these points, and therefore can be conveniently used 
to compute the Coulomb interaction between ions and/or electrons.
We note that the use of flat tori has been suggested in the solid state, in order to 
treat periodic boundary conditions, in the so-called cyclic-cluster approach
\cite{thomas_bredow_development_2001}.
%
Finally, we would like to point out that a torus formalism 
has also been suggested by Mamode in a similar context, in order to compute the 
Madelung constant of hypercubic crystals of any dimension, through the solution of the Poisson
equation in a finite space.
\cite{mamode_fundamental_2014,mamode_computation_2017}

In this work we apply our formalism to the case of crystals whose generating
vectors form an orthogonal system (but not necessarily an orthonormal one).
We note that, for practical computations, orthogonal supercells can fit
any type of unit cell to every arbitrary precision, provided sufficiently large supercells are considered.

This article is organized as follows.
In section 2 the methods used to compute the Madelung sums are briefly recalled.
In Section 3, a general presentation of the Clifford torus is given, and the two distances that
most naturally arise on it are discussed.
In Section 4 we define the Clifford supercell, and show how a two-point sum, such as as the Coulomb energy, 
over this lattice reduces, in the case of an orthogonal torus, 
to a single-point sum, because of the translational invariance of the system.
In Section 5 the computational details are presented, and the results obtained on 
1, 2, and 3-dimensional cubic systems are discussed.
The Madelung constants obtained by using finite supercells can be extrapolated to the infinite size system
in order to increase the accuracy of the result, as discussed in Section 6.
Finally, in Section 7 we draw our conclusions.

\section{Lattice Sums}

We briefly recall here some of the procedures that are used 
to compute the Madelung constant in ionic crystals.
This problem was first addressed by the German physicist Erwin Madelung \cite{madelung_e_electric_1918},
and the resulting series are therefore often called Madelung sums.
There is a conceptual difficulty associated to these calculations, 
and this is due to the fact that, because of the long-range nature of the Coulomb potential,
the resulting series are conditionally convergent, even in the case of 1-D systems.
For this reason, special care must be taken to perform the sums,
since the order of summation influences the result.

Mathematically, Madelung sums are a special case of lattice sums.
Let us consider a lattice $L$ in an $n$-dimensional space, 
and an infinite set ${\cal A}_L$ of real numbers defined on the lattice, 
${\cal A}_L  =  \{{a}_l\}$, with $l \in L$.
The lattice sum is defined as $S  =  \sum_{l\in L}a_l$.
Lattice sums became important in physics at the end of the 19-th century,
mainly in the theory of crystallography, and subsequently became a separate branch of mathematics.
Lattice sums can be performed either by direct summations or indirectly,
by integral transformations.
In the direct-summation methods, the different terms to be summed are suitably
regrouped into sets of increasing size (supercells), 
until all the terms of the sequence are eventually considered.
In doing this operation, the charges are sometimes modified, in order to ensure
the neutrality of the supercell.
The convergence of the direct procedure depends crucially on the shape and size of the supercells.
In particular, in order to achieve good convergence properties, it is important that
the total charge and the lowest multipole moments of the supercell vanish
\cite{coogan_simple_1967,gelle_fast_2008}.
Two early methods for the direct summation of Madelung sums have been proposed
by Evjen \cite{evjen_stability_1932} and sometimes later by H{\o}jendahl \cite{k_hojendahl_no_1938}.
These methods use fractional charges in order to ensure convergence.
In fact, it has been shown that neutrality of the supercell is required to accelerate 
convergence, and sometimes even to ensure convergence at all.
These approaches turns out to be a special case of the charge-renormalization techniques
\cite{sousa_madelung_1993,derenzo_determining_2000}.
It has also been shown that the rate of convergence is directly related to the number of
vanishing multipolar moments in the unit cell.
In this way, by imposing a number of vanishing moments in the cell, exponential 
convergence can be achieved \cite{gelle_fast_2008}.

The first and most commonly used integral-transformation
method, on the other hand, was proposed by Ewald \cite{ewald_berechnung_1921},
and several related methods have been proposed \cite{nijboer_calculation_1957}
(see, for instance, reference \cite{glasser_lattice_1980} 
for a review article with an extensive bibliography).
For a modern and comprehensive approach to the problem of lattice sums, see the excellent book by
J.M. Borwein, M.L. Glasser, R.C. McPhedran, J.G. Wan and I.J. Zucker,
``Lattice Sums Then and Now''  \cite{borwein_lattice_2013}.
This work follows a very mathematically oriented approach, and considers the problem of
giving a well defined and unambiguous definition of the conditionally convergent series
encountered in the Madelung formalism.
See also Ref.~\cite{borwein_convergence_1985}, and
the seminal paper of Glasser and Zucker~\cite{glasser_lattice_1980} .
The concept of analytic continuation of a complex function, in order 
to provide a basis for an unambiguous mathematical definition of Madelung's constant,
was introduced by Borwein, Borwein and Taylor\cite{borwein_convergence_1985}, who applied
this technique to the 2-D and 3-D calculation of the NaCl constants. 
It should be noticed, in fact, that the NaCl-type lattices 
(in one (1-D), two (2-D), or three dimensions (3-D))
are by far the most frequently considered, 
and a large number of papers are devoted to this subject \cite{borwein_convergence_1998}.
The NaCl Madelung sums are known with an extremely high accuracy (in the 1-D case, actually,
an analytical expression can be easily derived).
The Madelung constants for different crystals, on the other hand, are usually
computed with a much smaller accuracy.
Since in this work we concentrate ourselves on the convergence of the sums to a common limit
if cells of different size are used, we restrict our investigation to NaCl-like crystals.
In a previous paper, several different types of structures have been studied.~\cite{Tavernier_2020}

\section{Clifford Tori}

A Clifford torus is a closed border-less flat surface, 
first proposed and discussed by William Kingdon Clifford in 1873.
Its particularity is that it has a finite extension.
It is not possible to build a non-trivial Clifford torus in the ordinary 3-D real space, 
since the minimum
number of dimensions that permit to embed a Clifford torus is four (and six for a generalized 3-D Clifford torus).
Alternatively, one can work in a space having the same dimensionality of the torus, but which is complex.
Because of its clear and synthetic style, we report here a part of the introduction of
Klaus Volkert's historical article on the Bulletin of the Manifold Atlas: \cite{volkert_2013}
\cite{mcintosh_clifford_1999}
\noindent
{\em Clifford-Klein space forms entered the history of mathematics in 1873 during a
talk which was delivered by W. K. Clifford at the meeting of the British Association
for the Advancement of Sciences (Bradford, in September 1873) and via an article
he published in June 1873 \cite{clifford_preliminary_1871}. The title of Clifford's talk was ``On a surface of zero curvature and finite extension'', 
the proceedings of the meeting only provide this title. But we know a bit more about it from F. Klein 
who attended Clifford's talk and who described it on several occasions (for example\cite{klein_zur_1890}). 
In the context of elliptic geometry - which Clifford conceived in Klein's way as the
geometry of the part of projective space limited by a purely imaginary quartic -
Clifford described a closed surface which is locally flat, the today so-called Clifford
surface (this name was introduced by Klein \cite{klein_zur_1890}). This surface is constructed
by using (today so called) Clifford parallels; Bianchi later provided a description
by moving a circle along an elliptic straight line in such a way that it is always
orthogonal to the straight line \cite{bianchi_sulle_1896}. So Clifford's surface is the analogue
of a cylinder; but - since it closed - it is often called a torus. To show its local flatness,
Clifford used a dissection into parallelograms which is defined in a natural way by
the two sets of parallels (or generators) contained in the surface. By considering the
angles of such a parallelogram one sees that their sum equals four right angles so it
is a common flat parallelogram. Clifford concluded: ``The geometry of this surface
is the same as that of a finite parallelogram whose opposite sides are regarded as
identical'' \cite{clifford_preliminary_1871}. 
This is a very early occurrence of the identification scheme for
the torus!}

Besides the topology of the supercell, the second key ingredient of the formalism is the 
definition of a distance between two arbitrary points in the supercell.
This is a very important point: the Clifford torus being a multiply-connected manifold, 
an infinite number of different geodesics are possible to connect two points.
This means that the definition of the distance between to points requires some care.
In particular, it is clear that a unique distance between to points must be defined, 
in order to obtain a unique, well-defined inter-particle potential.
One could choose the length of the shortest one among the infinite number of paths as being
the distance between the two points.
We will call this distance the geodesic distance.
However, as soon as the points move on the torus, the shortest path switches from one geodesic to a different one.
This induces a jump in the derivative of the distance with respect to the position, 
and a similar discontinuity will be shown by the derivative of any potential which is a function of the inter-particle distance.
As a result, the force, which is the gradient of the potential, becomes a discontinuous
function of the particle positions, which is clearly a non-physical result.
For this reason, the geodesic Riemannian distance between two points does not seem to be a good candidate to be the
distance that is introduced into the Coulomb law and compute the potential between two charges.
However, as anticipated in the Introduction, 
there is a second metric that is also naturally defined for these manifolds,
and this is the metric of the embedding Euclidean space which contains the torus.
We will call the corresponding distance the Euclidean distance, and we will use precisely this distance in order to define the Coulomb potential of the system.

\begin{figure*}[!htb]
\begin{center}
\hspace{0mm}{{{
 \includegraphics[height=8cm,angle=0]{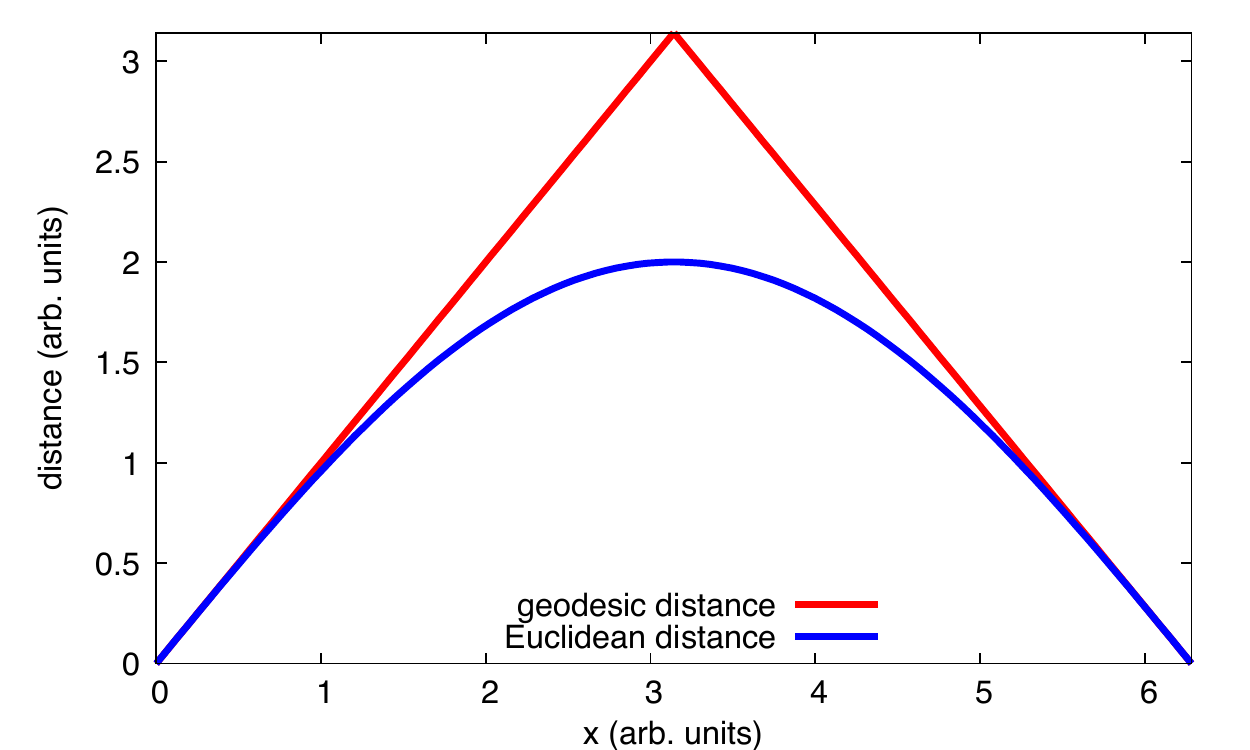}
}}}
\caption{The geodesic and Euclidean distance between two points on a circle of length $2\pi$ with one point fixed at $x=0$.}
\label{Figure1}
\end{center}
\end{figure*}

\begin{figure*}[!htb]
\begin{center}
\hspace{0mm}{{{
 \includegraphics[height=8cm,angle=0]{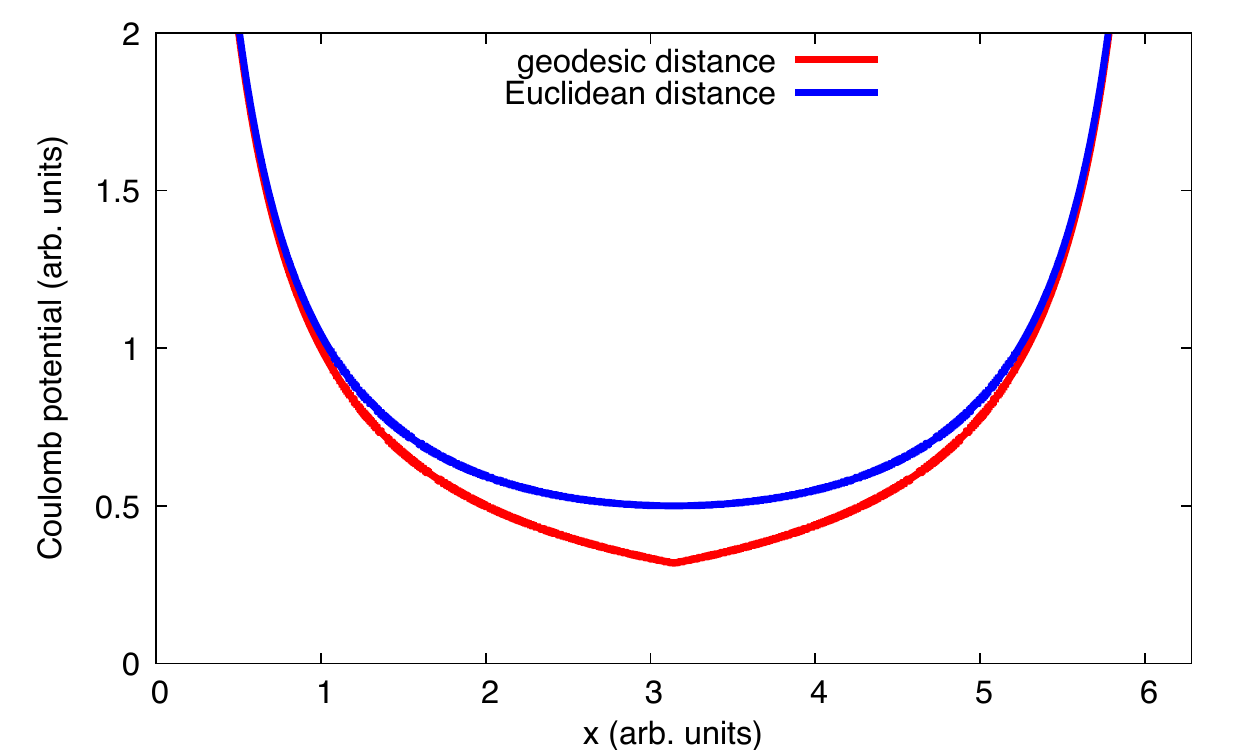}
}}}
\caption{The Coulomb potentials between two equal charges calculated using the geodesic and Euclidean distances in Fig.~\ref{Figure1}.}
\label{Figure2}
\end{center}
\end{figure*}

We illustrate these two different definitions of the distance between two points in the case of 
a 1-D Clifford torus, which is isomorphous to a circle.
In Fig.~\ref{Figure1} we report the geodesic and the Euclidean distances between two points on a 1-D torus of length $2\pi$ as a function of $x$ , 
which is the position of a point with respect to another point fixed at $x=0$.
We observe that the geodesic distance has a cusp at $x=\pi$.
Instead the Euclidean distance is a smooth function of $x$ everywhere.
In Fig.~\ref{Figure2} we report  the Coulomb potential computed using the two distances.
The potential obtained from the geodesic distance has a cusp at $x=\pi$ and, therefore, its derivative is discontinuous at that point.
The potential obtained from the Euclidean distance, on the other hand, is a unique and
differentiable function of $x$.
(except, of course, for the singularity that we have when the two charges coalesce, that is a feature common to any Coulomb potential).
We note that the Euclidean distance between two points is always shorter than the corresponding geodesic distance.
Therefore, for a given pair of points along the circle, the Coulomb potential computed by using the Euclidean distance is invariably larger, in absolute value, than that obtained with the geodesic distance.


The Euclidean distance on a Clifford torus tends, in the limit of a very large size, 
to the distance on an ordinary flat space having the same dimensionality.
%
%
For instance, for a 1-D Clifford torus of length $L$ the distance between two points whose coordinates differ by $x$ is given by 
\begin{equation}
    d_L(x)  =  \frac{L}{2\pi}\sqrt{2-2\cos\Bigl(\frac{2\pi x}{L}\Bigr)} .
\end{equation} 
If $x\in[0,2\pi]$, this expression can be rewritten as
\begin{equation}
    d_L(x)  =  \frac{L}{\pi}\sin\Bigl(\frac{\pi x}{L}\Bigr) .
\end{equation} 
It follows that, for a {\em fixed} value of $x$, we have
\begin{equation}
    \lim_{L\rightarrow \infty} d_L(x)  =  x .
\end{equation}
This means that the Euclidean distance between two points on the torus tends, under these circumstances, to the geodesic distance.
Therefore, the CSC Madelung constants will be expected to converge,
in the limit of very large supercells, to the ordinary ESC results, provided that the latter converge,
although we do not have a formal proof of this fact.
The main difficulty to formulate such a proof is that, unlike in the Evjen method, 
in our approach the Euclidean distance between two points depends on the size of the supercell.


\section{The Clifford supercell }

We consider a Bravais lattice in $n$ dimensions, with $n=1$, $2$, or $3$.
Notice that, although the value of $n$ for real systems is comprised between $1$ and $3$, this is not mandatory.
Let ${\bf v}_j$ be the generator vectors of a {\em unit} cell 
(not necessarily a {\em primitive unit} cell), and let $v_j=\|{\bf v}_j\|$.
A generic point $ {\bf u}_{\alpha_1...\alpha_n} $ belonging to the unit cell will be given by the vector
\begin{equation}
{\bf u}_{\alpha_1...\alpha_n}  =  \sum_{j=1}^n \alpha_j {\bf v}_j
\end{equation}
where the real parameters $\alpha_j$ verify the relation $0 \le \alpha_j < 1$.

Given a set of $n$ positive integers $K_j$, 
we define the Euclidean supercell (ESC) of sides ${\bf V}_1,...,{\bf V}_n$
as the parallelepiped generated by the $n$ vectors in $\mathbb{R}^n$
\begin{equation}
{\bf V}_{j} = K_j {\bf v}_j ,
\end{equation}
with $K_j \in \mathbb{N}$.
If the integers $K_j$ do not depend on $j$, {\em i.e.}, $K_j=K \; \forall \, j$, we have an equilateral supercell.
In the following discussion we will restrict ourselves to orthogonal supercells, i.e.,
\begin{equation}
{\bf v}_i \cdot {\bf v}_j =  \| {\bf v}_i \|  \| {\bf v}_j \| \delta_{ij},
\end{equation}
The ESC consists in a total of $K_1...K_n  =  \prod_{j=1}^n K_j$ replicas of the unit cell, 
$K_j$ for each one of the $n$ space directions $ {\bf v}_j $.
A generic lattice point in the ESC will be obtained by translating 
the lattice vector $ {\bf u}_{\alpha_1...\alpha_n} $, 
which is in the unit cell, by lattice vectors $k_j {\bf v}_j$ in each of the $j$ space directions,
where $k_j$ are integer numbers comprised between zero and $K_j-1$:
$0 \le k_j < K_j$.
This produces a new vector ${\bf W}_{\alpha_1...\alpha_n}$ given by
\begin{equation}
{\bf W}_{\alpha_1...\alpha_n}  =  {\bf u}_{\alpha_1...\alpha_n}  +  \sum_{j=1}^n k_j {\bf v}_j .
\end{equation}
Therefore, in term of the unit-cell generators ${\bf v}_j$, 
the vector ${\bf W}$ can be expressed as
\begin{equation}
{\bf W}_{x_1...x_n}
 =  \sum_{j=1}^n (\alpha_j + k_j) {\bf v}_j
\; \equiv \; \sum_{j=1}^n x_j {\bf v}_j ,
\end{equation}
where we defined the new variables $x_j \, \equiv \, \alpha_j + k_j$.
We note that the sides of the ESC will have length $ K_j v_j$.
In the following, if not strictly necessary, 
we will drop the subscripts $x_1,...,x_n$ for economy of notation, and write
simply $ {\bf W} $ instead of $ {\bf W}_{x_1...x_n} $.

In a completely similar way, we define the Clifford supercell (CSC) as the Clifford torus in $\mathbb{C}^n$
associated to the ESC, obtained by joining the opposite edges of the corresponding ESC. 
Since we are now in a complex field, we must distinguish between a complex vector 
$|{\bf W}\rangle$ and its adjoint $\langle{\bf W}|$.
We take as generators $|{\bf v}_j\rangle$ of the CSC the same generators of the ESC, 
$|{\bf v}_j\rangle \, \equiv \, {\bf v}_j$.
As for the real case, we assume the orthogonality property of the basis vectors,
\begin{equation}
\langle {\bf v}_i | {\bf v}_j\rangle  =  \| {\bf v}_i \|  \| {\bf v}_j \| \delta_{ij} .
\end{equation}
A generic point in the CSC is given by
\begin{equation}
|{\bf W}_{\theta_1...\theta_n}\rangle  =   \sum_{j=1}^n 
\frac{K_j}{2\pi} \exp(i \theta_j) |{\bf v}_j\rangle ,
\end{equation}
where $\theta_j  =  2\pi x_j/ K_j $.
The factor $ \frac{K_j}{2\pi}$ has been chosen in such a way that the length of circle with
radius $ \frac{K_j v_j}{2\pi}$
coincides with the length of the corresponding side of the ESC, i.e.,  $K_j v_j$. 
Since the ESC and the corresponding CSC are built by using the same unit vectors $ {\bf v}_j$, 
this means that the two cells are locally isometric.
In the following, we will drop the subscripts $\theta_1,...,\theta_n$ for economy of notation, and write
for simplicity $ |{\bf W}\rangle $ instead of $ |{\bf W}_{\theta_1...\theta_n}\rangle $.

We come now to the problem of computing the classical Coulomb energy of a set of charges distributed in a CSC.
As discussed in the introduction, the Riemann distance between two points is not uniquely defined 
since it is a multi-valued function, and therefore it is 
not suitable to define the distance-dependent Coulomb potential.
Instead, we assume as the norm of a vector in the CSC the usual Euclidean norm in 
the embedding space, $\mathbb{C}^n$.
Notice, however, that since the CSC manifold is not a linear space, 
the structure of the CSC is not that of a Banach space.
Finally, we notice that, since $\mathbb{C}$ is isomorphous to $\mathbb{R}^2$, the CSC can be also
seen as a set of points belonging to a manifold in $\mathbb{R}^{2n}$.
Let us now discuss in detail the cases of the two and three dimensional Clifford supercells.

\subsection{the 2-D Clifford supercell}

In order to compute Madelung sums,
we have to consider the sum over a set of points in the CSC of a function that depends on the distance
between the points.
Let us consider two points, $A$ and $B$, in the CSC:
\begin{align}
|{\bf W}^A\rangle  &=  \frac{K_1}{2\pi} \, \exp(i \theta_1^A) \, |{\bf v}_1\rangle + 
\frac{K_2}{2\pi} \, \exp(i \theta_2^A) \, |{\bf v}_2\rangle
\\
|{\bf W}^B\rangle  &=  \frac{K_1}{2\pi} \, \exp(i \theta_1^B) \, |{\bf v}_1\rangle + 
\frac{K_2}{2\pi} \, \exp(i \theta_2^B) \, |{\bf v}_2\rangle .
\end{align}
Since the CSC is embedded in $\mathbb{C}^n$, we define the Euclidean distance 
between the two points $A$ and $B$ as the usual norm, in $\mathbb{C}^n$, 
of the difference $|{\bf W}^{AB} \rangle$ of the two corresponding position vectors.
This difference $|{\bf W}^{AB}\rangle$ between the two vectors $|{\bf W}^B\rangle$ and $|{\bf W}^A\rangle$ is given by

\begin{align}
|{\bf W}^{AB}\rangle  =  |{\bf W}^{B}\rangle - |{\bf W}^{A}\rangle & = 
\nonumber
\frac{K_1}{2\pi} \, \bigl(\exp(i \theta_1^B) - \exp(i \theta_1^A) \bigl) \, |{\bf v}_1\rangle  
\\ &+ 
\frac{K_2}{2\pi} \, \bigl(\exp(i \theta_2^B) - \exp(i \theta_2^A) \bigl) \, |{\bf v}_2\rangle
\end{align}

Let us now consider a double lattice sum
\begin{equation}
F  =  \sum_{A} \sum_{B} \, {\cal F} \left( \|{\bf W}^{AB} \|   \right) 
\end{equation}
where both $A$ and $B$ run over the same set of points in the CSC, 
and $\cal F$ is an arbitrary real function.
Notice that the two sets, $A$ and $B$, can coincide.
We assume that the sums in the above equation are well-defined.
We have that
\begin{align}
\nonumber
\|{\bf W}^{AB} \|^2  &= 
\frac{K_1^2}{4\pi^2}
\bigl(e^{-i\theta_1^B}-e^{-i\theta_1^A}\bigr)
\bigl(e^{i\theta_1^B}-e^{i\theta_1^A}\bigr) 
\langle {\bf v}_1|{\bf v}_1\rangle
\\ &+ \frac{K_2^2}{4\pi^2} \, \bigl(e^{-i\theta_2^B}-e^{-i\theta_2^A}\bigr)\bigl(e^{i\theta_2^B}-e^{i\theta_2^A}\bigr) 
\nonumber
\langle{\bf v}_2|{\bf v}_2\rangle
\\& + \frac{K_1 K_2}{4\pi^2} \, \bigl(e^{-i\theta_1^B}-e^{-i\theta_1^A}\bigr)\bigl(e^{i\theta_2^B}-e^{i\theta_2^A}\bigr) 
\nonumber
\langle{\bf v}_1|{\bf v}_2\rangle
\\&+ \frac{K_1 K_2}{4\pi^2} \, \bigl(e^{-i\theta_2^B}-e^{-i\theta_2^A}\bigr)\bigl(e^{i\theta_1^B}-e^{i\theta_1^A}\bigr) 
\langle{\bf v}_2|{\bf v}_1\rangle .
\end{align}

By taking into account that we assumed to deal with an orthogonal CSC, 
i.e., $\langle{\bf v}_2|{\bf v}_1\rangle  =  0 $,
the above expression can be rewritten as
\begin{equation}
\|{\bf W}^{AB} \|^2  = 
\frac{K_1^2}{2\pi^2} \, \bigl(1-\cos(\theta_1^B-\theta_1^A)\bigr) \|{\bf v}_1\|^2 + \frac{K_2^2}{2\pi^2} \, \bigl(1-\cos(\theta_2^B-\theta_2^A)\bigr) \|{\bf v}_2\|^2 .
\end{equation}
It is convenient at this point to use the explicit expressions for the angles $\theta_j^X$.
We thus obtain
\begin{align}
\|{\bf W}^{AB} \|^2  &= 
\nonumber
\frac{K_1^2}{2\pi^2} \, \Bigl(1-\cos\bigl(\frac{\alpha_1^B-\alpha_1^A+k_1^B-k_1^A}{K_1}\bigr)\Bigr) 
\|{\bf v}_1\|^2
\\ &+
\frac{K_2^2}{2\pi^2} \, \Bigl(1-\cos\bigl(\frac{\alpha_2^B-\alpha_2^A+k_2^B-k_2^A}{K_2}\bigr)\Bigr) 
\|{\bf v}_2\|^2 .
\end{align}

The above expression could be further expanded by using the expressions for the cosine of a sum
of angles, but this is not really needed.
Indeed, our intent here is simply to show that the four-index sums, involved in the 
evaluation of the Coulomb energy, break down to much simpler two-index sums.
To this aim, we compute the sum $F$, expressed as
\begin{align}
F  &= 
\nonumber
\sum_{A \in u.c}\sum_{B \in u.c}
\sum_{k_1^A=0}^{K_1-1}
\sum_{k_2^A=0}^{K_2-1}
\sum_{k_1^B=0}^{K_1-1}
\sum_{k_2^B=0}^{K_2-1}
 {\cal F} \, \Bigl[\frac{K_1^2}{2\pi^2} \bigl(1-\cos(\frac{\alpha_1^B-\alpha_1^A+k_1^B-k_1^A}{K_1})\bigr) 
\|{\bf v}_1\|^2
\\ &+  
\frac{K_2^2}{2\pi^2} \bigl(1-\cos(\frac{\alpha_2^B-\alpha_2^A+k_2^B-k_2^A}{K_2})\bigr) \|{\bf v}_2\|^2 \Bigr],
\end{align}
where the sum over $A$ and $B$ is now restricted to just the unit cell.
We now define the new variables $k_1 \equiv k_1^B-k_1^A$ and $k_2 \equiv k_2^B-k_2^A$.
Because of the periodicity of the the CSC, the indices $k_1$ and $k_2$ are associated to 
the cells $(k_1^B-k_1^A \mod \, K_1)$ and  $(k_2^B-k_2^A \mod \, K_2)$, respectively.
In this way, and since the sums are over all the possible values of $k_1$ and $k_2$, we can rewrite $F$ as
\begin{align}
F &=
\nonumber
K_1 K_2
\sum_{A \in u.c}\sum_{B \in u.c}
\sum_{k_1=0}^{K_1-1}
\sum_{k_2=0}^{K_2-1}
{\cal F} \Bigg[\frac{K_1^2}{2\pi^2} \bigl(1-\cos(\frac{\alpha_1^B-\alpha_1^A+k_1}{K_1})\bigr) \|{\bf v}_1\|^2
\\ & +  
\frac{K_2^2}{2\pi^2} \bigl(1-1\cos(\frac{\alpha_2^B-\alpha_2^A+k_2}{K_2})\bigr) \|{\bf v}_2\|^2 \Bigg].
\end{align}
In this way, the original four-index summation yielding $F$ reduces to a two-index summation.
This fact implies a huge computational saving.

\subsection{The 3-D Clifford supercell}

We briefly report here the analogous result for the 3-D Clifford supercell.
The two points $A$ and $B$ will be given now by
\begin{align}
|{\bf W}^A \rangle  &=  \frac{K_1}{2\pi} \, 
\exp(i \theta_1^A) |{\bf v}_1 \rangle + 
\frac{K_2}{2\pi} \, \exp(i \theta_2^A) |{\bf v}_2 \rangle  +
\frac{K_3}{2\pi} \, \exp(i \theta_3^A) |{\bf v}_3 \rangle
\\
| {\bf W}^B \rangle  &=  \frac{K_1}{2\pi} \, 
\exp(i \theta_1^B) |{\bf v}_1 \rangle + 
\frac{K_2}{2\pi} \, \exp(i \theta_2^B) |{\bf v}_2 \rangle  +
\frac{K_3}{2\pi} \, \exp(i \theta_3^B) |{\bf v}_3 \rangle
,
\end{align}
and the lattice sum $F$ becomes
\begin{align}
F  &= 
\nonumber
\sum_{A \in u.c}\sum_{B \in u.c}
\sum_{k_1^A=0}^{K_1-1}
\sum_{k_2^A=0}^{K_2-1}
\sum_{k_3^A=0}^{K_3-1}
\sum_{k_1^B=0}^{K_1-1}
\sum_{k_2^B=0}^{K_2-1}
\sum_{k_3^B=0}^{K_3-1}
{\cal F} \Bigl[\frac{K_1^2}{2\pi^2} \bigl(1-\cos(\frac{\alpha_1^B-\alpha_1^A+k_1^B-k_1^A}{k})\bigr) \|{\bf v}_1\|^2 \; +
\\ &+ 
\frac{K_2}{2\pi^2} \bigl(1-\cos(\frac{\alpha_2^B-\alpha_2^A+k_2^B-k_2^A}{k})\bigr) 
\|{\bf v}_2\|^2  +  \frac{K_3}{2\pi^2} \bigl(1-\cos(\frac{\alpha_3^B-\alpha_3^A+k_3^B-k_3^A}{k})\bigr) 
\|{\bf v}_3\|^2 \Bigr]
\end{align}
By applying the same arguments used for the two-dimensional case we arrive at
\begin{align}
\nonumber
F  &=  K_1 K_2 K_3
\sum_{A \in u.c}\sum_{B \in u.c}
\sum_{k_1=0}^{K_1-1}
\sum_{k_2=0}^{K_2-1}
\sum_{k_3=0}^{K_3-1}
\, {\cal F} \, \Bigl[\frac{K_1^2}{2\pi^2} \bigl(1-\cos(\frac{\alpha_1^B-\alpha_1^A+k_1}{K_1})\bigr) 
\|{\bf v}_1\|^2 +
\\ &+  
\frac{K_2^2}{2\pi^2} \, \bigl(1-\cos(\frac{\alpha_2^B-\alpha_2^A+k_2}{K_2})\bigr) \|{\bf v}_2\|^2
+\frac{K_3^2}{2\pi^2} \, \bigl(1-\cos(\frac{\alpha_3^B-\alpha_3^A+k_3}{K_3})\bigr) \|{\bf v}_3\|^2 \Bigr] \;
\end{align}

These results show that, in the case of orthogonal supercells, the sums over all the atom pairs reduce to much more manageable sums over the individual atoms only.
The translational invariance of the lattice points in ${\cal T}^n$ implies that corresponding
points in different unit cells ``see'' exactly the same environment. 
This property holds for any type and content of the unit cell, 
contrary to what happens for points on a sphere ${\cal S}^n$:
in fact, it is not possible to cover a sphere with a regular mesh
of points, except for a few special values of the number of points.
%
%
\subsection{The Madelung constant}
In this work the double lattice sum of interest is the cohesion energy $E_{coh}$ due to the Coulomb interactions of the ions in a crystal.
It is given by 
\begin{equation}
E_{coh} = \frac{1}{2 R_0}\sum_{A\in u.c.} z_A M_A,
\end{equation}
where
$z_A$ is the valency of ion $A$, $R_0$ is the nearest-neighbor distance, and $M_A$ is the Madelung constant which, in terms of the renormalized distance, is defined by~\cite{Tavernier_2020}
\begin{equation}
M_A = 
R_0 \sum_{B\in u.c.}\!\!\!\!{\vphantom{\sum}}' z_B \sum_{k_1=0}^{K_1-1} \cdots \sum_{k_n=0}^{K_n-1}
\Bigg[ \sum_{j=1}^n
\frac{K_j^2}{2\pi^2}
\bigg[1-
\cos\left(\frac{2\pi}{K_j}[\alpha_j^B - \alpha_j^A + k_j]\right)\bigg]\|{\bf v}_j\|^2
\Bigg]^{-1/2},
\label{Eqn:Madelung_CSC}
\end{equation}
where the prime indicates that $B$ runs over the positions of all ions in the unit cell except that of the atom $A$ when $k_i =0 \forall i$.
%
\section{Computational Details}
In this section we describe the computational techniques used to calculate the Madelung sums and present the obtained results.
We compare three different methods:
\begin{enumerate}
    \item 
    The plain sum over ESC of increasing size.
    This approach gives in general non-converging sums, and is added for completeness.
    \item
    The plain sum over ESC of increasing size but with border-weighted charges such that the ESC is neutral (Evjen Method).
    \item
    The sum over CSC with the embedding distance, which is the method proposed in the present work.
\end{enumerate}
We note that in method 1 the supercell is not charge neutral, 
and in most cases this method fails to converge to a unique value.
Also, Method 2 can give two different limit values, as in the well known cases of CsCl and ZnS.~\cite{Tavernier_2020} 

All results were obtained with a simple code of about a hundred lines that is freely available.~\cite{Madelung_code}
Since the numerical precision is a key issue in order to evaluate the large-size behaviour of the
Madelung constants, all our results have been obtained using quadruple precision, although double precision is usually more than enough for standard calculations.
Finally, in order to facilitate the reproduction of our results, we report for each calculation
with neutral supercells, i.e., for methods 2 and 3, the number of atoms in the supercell.

\subsection{NaCl}

NaCl is by far the most studied lattice, and a huge number of articles are available, not only for the 3-D
case, but also for the 1-D and 2-D structures.
Therefore, we consider here in detail all the three cases ($n=3$, but also $n=2$ and $n=1$),
although formally, for instance, in the $n=2$ case there is no difference between NaCl and CsCl.
In the figures of this section we will report the values of the
Madelung constant $M(K)$ as a function of $1/K^p$, where $p$ is an integer chosen such that 
$M(K)$ as a function of $1/K^p$ is linear.
This functional dependence is particularly suitable in order to extrapolate the $M(K)$ value 
to the limit $K \rightarrow \infty$, that gives the Madelung constant for the bulk.

\subsubsection{1-D ``NaCl''}

This case admits a simple analytical solution, 
since the alternating harmonic series is known to converges to $\ln 2$ \cite{hudelson_proof_2010}.
Therefore, the 1-D NaCl Madelung sum is given by minus twice this value, $-2\ln 2$.
Because of the fact that the exact value can be easily computed to any given accuracy, 
this structure permits 
a straightforward evaluation of the absolute error associated to the different methods.

\subsubsection{2-D ``NaCl''}

\begin{figure*}[!htb]
\begin{center}
\hspace{0mm}{{{
 \includegraphics[height=8cm,angle=0]{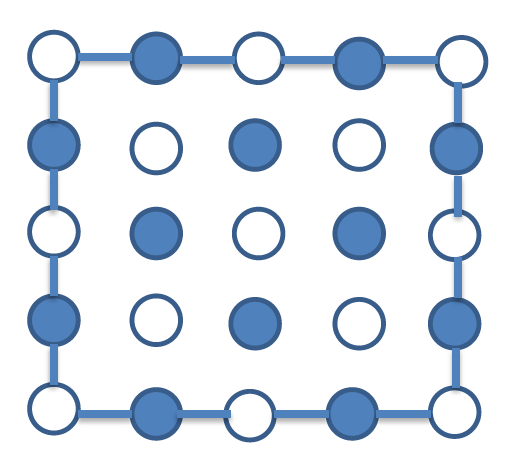}
}}}
\caption{An illustration of the structure of 2-D ``NaCl'': standard geometry}
\label{Figure3}
\end{center}
\end{figure*}

\begin{figure*}[!htb]
\begin{center}
\hspace{0mm}{{{
 \includegraphics[height=8cm,angle=0]{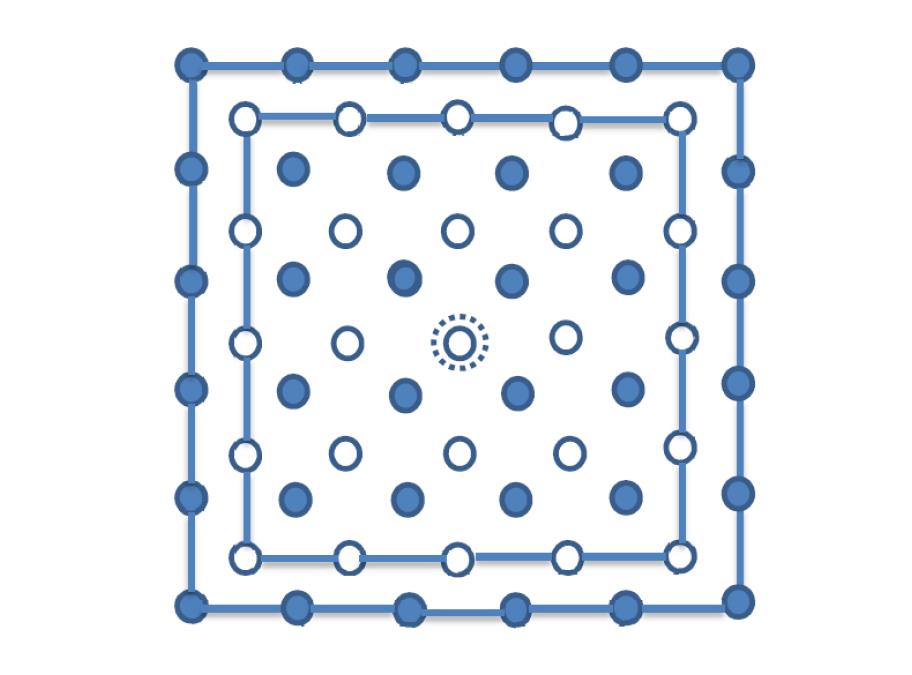}
}}}
\caption{An illustration of the structure of 2-D ``NaCl'': tilted geometry}
\label{Figure4}
\end{center}
\end{figure*}

We study two types of cells: one that can be seen as composed of a single sheet of 3-D NaCl, 
and for which the atoms of the same type are placed over straight lines parallel to the $x$ and $y$ axes (see Fig.~\ref{Figure3})
We will call refer to this geometry as the standard geometry.
A second choice is possible, in which the standard geometry is rotated by $45^{\circ}$  (see Fig.~\ref{Figure4})..
We will call refer to this geometry as the tilted geometry.
Although these two structures are obviously equivalent, the different methods have very different convergence properties, as discussed in detail in the next section.

\subsubsection{3-D NaCl}

The Madelung constant for 3-D NaCl crystal has been evaluated with many tens of digits 
of accuracy. 
Its fifteen-digits approximate value is $-1.74756459463318$ \cite{oeis_NaCl}.
We computed the sum by using CSC having different sizes:
equilateral supercells of the type $(K, \, K, \, K)$, indicated with (1,1,1), 
as well as two types of non-equilateral 
supercells, namely $(K_1=K, \, K_2=K, \, K_3=4K)$, denoted $(1,1,4)$, and
$(K_1=K, \, K_2=2K, \, K_3=4K)$, denoted $(1,2,4)$.

\subsection{Extrapolation of CSC results}

The calculation of the Madelung constant of a single supercell, even of large size, do not
give an approximation of the exact value, corresponding to the infinite system, with more than five or six digits.
It is possible, however, to extrapolate the results for the finite-size CSC in order to evaluate the Madelung constant of the infinite crystal.
In order to perform the extrapolation, we studied the large-size behavior of $M(K)$.
It turns out that the CSC Madelung constant follows the following simple inverse power law as a function of the system size for large $K$,
\begin{equation}
    M(K)  =  M_\infty + A K^{-2}.
    \label{Eqn:extrapolation}
\end{equation}
which contains only two parameters $M_\infty$ and $A $.
In practice, we used the two largest values of $K$ to perform 
the extrapolation and thus obtain the Madelung constant, $M_\infty$, of the infinite system.
The extrapolated Madelung constants obtained in such a way are reported in the tables of the next section.

\section{Results and Discussion: the Madelung constants }

\begin{figure*}[!htb]
\begin{center}
\hspace{0mm}{{{
 \includegraphics[height=8cm,angle=0]{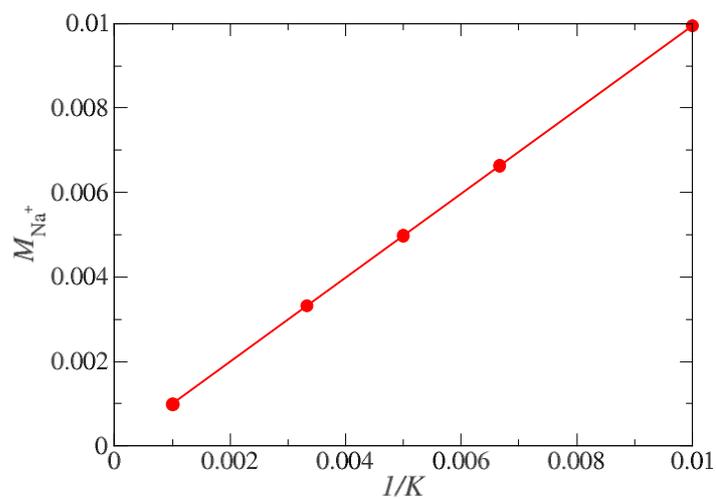}
}}}
\caption{The error in the Na$^+$ Madelung constant of 1-D ``NaCl'' with respect to the exact value $-2\ln 2$ as a function of $1/K$ calculated from a plain sum in the ESC.}
\label{Figure5}
\end{center}
\end{figure*}

\begin{figure*}[!htb]
\begin{center}
\hspace{0mm}{{{
 \includegraphics[height=8cm,angle=0]{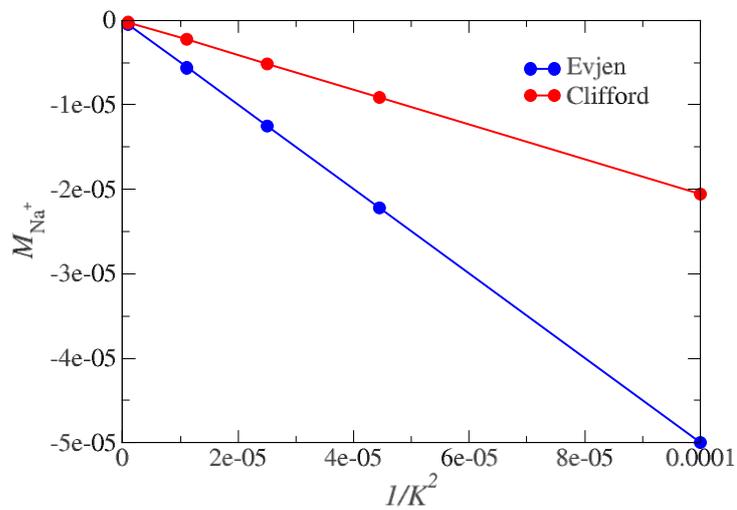}
}}}
\caption{The error in the Na$^+$ Madelung constant of 1-D ``NaCl'' with respect to the exact value $-2\ln 2$ as a function of $1/K^2$ calculated with Evjen's approach and with our CSC approach.}
\label{Figure6}
\end{center}
\end{figure*}

In Table 1, the results for the 1-D ``NaCl'' linear structure are presented.
In the table, the absolute errors with respect to the -$2\ln{2}$ exact value are shown.
The convergence of ESC method is extremely slow (in $1/K$), 
due to the the presence of non-zero charges in the supercells (see also Fig.~\ref{Figure5}).
Evjen's sum has a much faster convergence ($1/K^2$), but shows an alternating behavior.
Our CSC approach, on the other hand, is monotonically and rapidly 
(also as $1/K^2$) converging to the exact value.
These results are shown if Fig.~\ref{Figure6}, where both Evjen's sums (even values of K only)
and the CSC results are shown.

\begin{table*}
\caption{The error in the Na$^+$ Madelung constant of 1-D ``NaCl'' with respect to the exact value $-2\ln 2$.
The extrapolated CSC Madelung constant has been obtained according to Eq.~\eqref{Eqn:extrapolation} using the values of $K$ corresponding to the two largest supercells.}
\begin{center}
\begin{tabular}{rrrcccccl}
\hline
\\
      SC size ($K$) & & atoms & &  ESC & &  Evjen ESC & &  CSC   \\
\\
\hline
\\
           100 & &  200 & &  9.950002e-3  & &  -4.999750e-5  & &      -2.056137e-5  \\
\\
           101 & &  202 & & -9.851977e-3  & &   4.901240e-5  & &      -2.015624e-5  \\
\\
           102 & &  204 & &  9.755865e-3  & &  -4.805612e-5  & &      -1.976296e-5  \\
\\
           103 & &  206 & & -9.661610e-3  & &   4.712757e-5  & &      -1.938108e-5  \\
\\
           300 & &  600 & &  3.327777e-3  & &  -5.555524e-6  & &      -2.284626e-6  \\
\\
          1000 & & 2000 & &  9.995000e-4  & &  -4.999997e-7  & &      -2.056167e-7  \\
\\
\hline
\\
  extrap. & & & & & & & & -3.288287e-13 \\
\\
\hline
\end{tabular}
\end{center}
\end{table*}

\begin{figure*}[!htb]
\begin{center}
\hspace{0mm}{{{
 \includegraphics[height=8cm,angle=0]{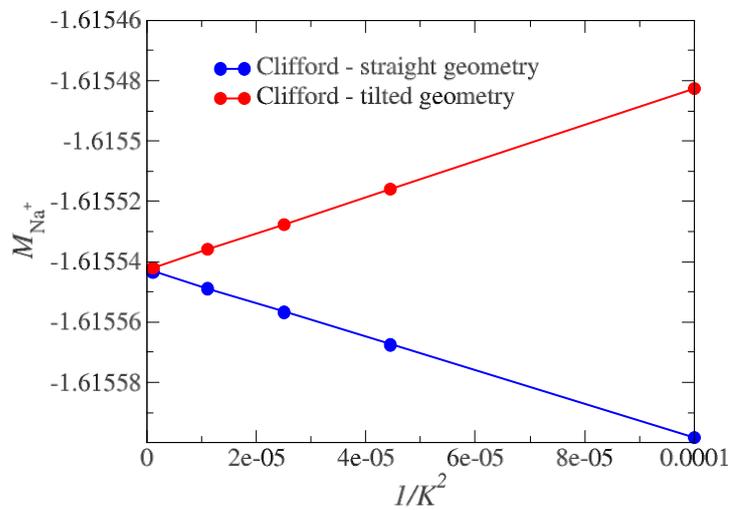}
}}}
\caption{The Na$^+$ Madelung constant of 2-D ``NaCl''  as a function of $1/K^2$ calculated with our CSC approach for the standard and tilted geometries.}
\label{Figure7}
\end{center}
\end{figure*}

In Tables 2 and 3, the results for the 2-D ``NaCl'' structure are presented.
We performed the calculations on both right (Table 2) and tilted (Table 3) cell structures.
In the case of the right structure, the plain summation using the ESC appears to be monotonically
slow ($1/K$) converging to the exact results. 
Evjen's sum is also convergent, with a much faster $1/K^3$ behavior.
Our CSC method has an intermediate $1/K^2$ behavior.
Things are remarkably different when the tilted cells are used.
In this case, the plain ESC method appears to converge to two different limits,
accordingly to the parity of $K$.
Evjen ESC converges to the exact limit, but with an alternating behavior.
The CSC approach is monotonically convergent to the exact limit.
The quality of the CSC method has been assessed by comparing the extrapolated energy
of the right and tilted supercells.
By using a two-point extrapolation according to Eq.\eqref{Eqn:extrapolation} the two supercells give results that coincide up to $10^{-10}$.
The CSC results are also illustrated in Fig.~\ref{Figure7}.

\begin{table*}
\caption{The Na$^+$ Madelung constant of 2-D ``NaCl''.
The extrapolated CSC Madelung constant has been obtained according to Eq.~\eqref{Eqn:extrapolation} using the values of $K$ corresponding to the two largest supercells.}
\begin{center}
\begin{tabular}{rrrcccccl}
\hline
\\
      SC size ($K$) & & atoms & &  ESC & &  Evjen ESC & &  CSC   \\
\\
\hline
\\
           100 & & 40000  & & -1.6085067818  & &   -1.6155424941 & &    -1.6155983700      \\
\\
           101 & & 40804  & & -1.6085760997  & &   -1.6155424980 & &    -1.6155972717      \\
\\
           102 & & 41616  & & -1.6086440650  & &   -1.6155425017 & &    -1.6155962055      \\
\\
           103 & & 42436  & & -1.6087107169  & &   -1.6155425053 & &    -1.6155951703      \\
\\
          300 & & 360000  & & -1.6131895275  & &   -1.6155426218 & &    -1.6155488207      \\
\\
        1000 & & 4000000  & & -1.6148358733  & &   -1.6155426265 & &    -1.6155431841      \\
\\
\hline
\\
  extrap. & & & & & & & & -1.61554262671649 \\
\\
\hline
\end{tabular}
\end{center}
\end{table*}

\begin{table*}
\caption{The Na$^+$ Madelung constant of 2-D ``NaCl'' in the tilted geometry.
The extrapolated CSC Madelung constant has been obtained according to Eq.~\eqref{Eqn:extrapolation} using the values of $K$ corresponding to the two largest supercells.}
\begin{center}
\begin{tabular}{rrrcccccl}
\hline
\\
      SC size ($K$) & & atoms & &  ESC & &  Evjen ESC & &  CSC   \\
\\
\hline
\\
           100 & & 20000  & & -0.8773253336  & &   -1.6255422934 & &    -1.6154827314      \\
\\
           101 & & 20402  & & -4.1084112340  & &   -1.6056419601 & &    -1.6154839119      \\
\\
           102 & & 20808  & & -0.8773266088  & &   -1.6253462342 & &    -1.6154850579      \\
\\
           103 & & 21218  & & -4.1084124722  & &   -1.6058341938 & &    -1.6154861706      \\
\\
          300 & & 180000  & & -0.8773546424  & &   -1.6188759477 & &    -1.6155359735      \\
\\
        1000 & & 2000000  & & -0.8773580008  & &   -1.6165426263 & &    -1.6155420279      \\
\\
\hline
\\
  extrap. & & & & & & & & -1.61554262673346 \\
\\
\hline
\end{tabular}
\end{center}
\end{table*}

In Table 4, we report the 3-D NaCl Madelung constants computed through 
a set of increasing-size cubic supercells of different types.
The plain sum over larger and larger ESC (which are electrically charged)
is very slowly convergent, as $K^{-1}$, to the exact limit with a series 
of values alternating around the exact value, depending on the total charge of the supercell.
In the 3-D NaCl case, the Evjen's sums converge extremely fast, as $K^{-4}$.
Finally, the Clifford series converges monotonically as $K^{-2}$.
In the table we report also the extrapolated CSC value, which coincide with the
exact result up to the ninth decimal digit.

\begin{figure*}[!htb]
\begin{center}
\hspace{0mm}{{{
 \includegraphics[height=8cm,angle=0]{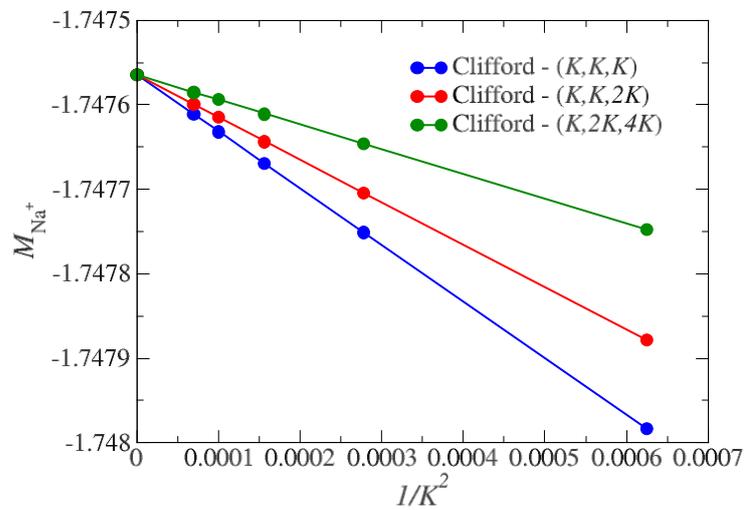}
}}}
\caption{The Na$^+$ Madelung constant as a function of $1/K^2$ calculated with our CSC approach for different shapes of the supercell.
The values at $K=0$ are obtained according to Eq.~\eqref{Eqn:extrapolation} using the values of $K$ corresponding to the two largest supercells.}
\label{Figure8}
\end{center}
\end{figure*}

\begin{table*}
\caption{The Na$^+$ Madelung constant of 3-D NaCl.
The extrapolated CSC Madelung constant has been obtained according to Eq.~\eqref{Eqn:extrapolation} using the values of $K$ corresponding to the two largest supercells.}
\begin{center}
\begin{tabular}{rrrcccccl}
\hline
\\
      SC size ($K$) & & atoms & &  ESC & &  Evjen ESC & &  CSC   \\
\\
\hline
\\
          40 & &  512000 & & -1.7333090325   & &    -1.7475646102    & &   -1.7479830134   \\
\\
          41 & &  551368 & & -1.7614766492   & &    -1.7475645804    & &   -1.7479628535   \\
\\
          42 & &  592704 & & -1.7339798824   & &    -1.7475646075    & &   -1.7479441161   \\
\\
          43 & &  636056 & & -1.7608370145   & &    -1.7475645829    & &   -1.7479266706   \\
\\
          60 & & 1728000 & & -1.7380216149   & &    -1.7475645977    & &   -1.7477505682   \\
\\
          80 & & 4096000 & & -1.7403925416   & &    -1.7475645956    & &   -1.7476692067   \\
\\
         100 & & 8000000 & & -1.7418198158   & &    -1.7475645950    & &   -1.7476315469   \\
\\
         120 & &13824000 & & -1.7427733060   & &    -1.7475645948    & &   -1.7476110895   \\
\\
\hline
\\
  extrap. & & $(K,K,K)$   & & & & & &  -1.74756459533123 \\
\\
          & & $(K,K,2K)$  & & & & & &  -1.74756459474737 \\
\\
          & & $(K,2K,4K)$ & & & & & &  -1.74756459440130 \\
\\
\hline
\\
  ref. \cite{oeis_NaCl}   & & & & & & & &  -1.74756459463318 \\
\\
\hline
\end{tabular}
\end{center}
\end{table*}
%
%
In order to assess the performance of the method when a generic, non-equilateral supercell is used, 
we compared the Madelung constant computed with the equilateral $(K,K,K)$ CSC with a $(K,K,2K)$ CSC and a $(K,2K,4K)$ CSC.
We extrapolated the Madelung constant by using, for all the three cases, the two
largest values of $K$, $K=100$ and $K=120$.
The results are reported in Table 4, and illustrated in Fig.~\ref{Figure8}. 
The three curves converge to a common limit with a $1/K^2$ law.
In the table we also report the extrapolated results for infinite-size systems.
Quite remarkably, the CSC extrapolated numerical results coincide with the ``exact''
value (taken from Ref.~\cite{oeis_NaCl}) up to ten decimal digits.

\section{Conclusions and Outlook}

In this article, we described a formalism suitable for the computation 
of the Madelung constants of ionic crystals via a real-space calculation of Madelung sums.
The general strategy of the proposed approach consists in transforming a large fragment
(a supercell) of a periodic system into a Clifford torus, and then redefining the distance between two
points by taking the Euclidean distance between these points in the embedding space of the torus.
In this way, a Madelung sum of an infinite periodic system is replaced by a sequence of sums over finite periodic systems, 
and the value for the infinite crystal is then obtained by extrapolating this sequence of energies to the infinite-size limit.
Moreover, our method is very simple and scales linearly as a function of the number of atoms.
Therefore, even the largest calculations performed in this work only take a few seconds on a laptop computer.

As a numerical application, we computed the Madelung constant of NaCl-type cubic ionic lattices in one, two and three dimensions. 
The obtained Madelung sum matches very well with the available reference data.
We used right supercells having different shapes, and we demonstrated that the value computed in this way does not depend 
on the shape of the parallelepiped supercell. 
This is very important for general (triclinic) crystals, since, by taking a supercell sufficiently large, any
triclinic supercell can be approximately fitted into a square parallelepiped within any chosen precision.

We note that our approach is general and not limited to the calculation of Madelung constants.
It is, for example, also useful in the study of Wigner localization ~\cite{Diaz-Marquez_2018,escobar_azor_wigner_2019} in periodic systems.
Exactly the same formalism as described in this work can be used to compute the energy of 
Wigner crystals~\cite{Wigner_1934}, the only difference being that the discrete sums must be replaced by 
integrals when the continuous jellium background charge is considered.~\cite{Alves_2021}
Our preliminary results on Wigner crystals indicate that the correct energy order of the different types of crystal structures
is recovered by using Clifford boundary conditions, both for 2-D and 3-D crystals.
Finally, it would be interesting to apply our approach to non-regular ionic systems.


\section*{Acknowledgements }

We wish to dedicate this article to our friend and colleague 
Fernand Spiegelman, of the Laboratoire de Chimie et Physique Quantiques in Toulouse, 
on occasion of his retirement.

We would like to thank prof. Richard E. Schwartz (Brown University) for helpful discussions.
This work was partly supported by the French
``Centre National de la Recherche Scientifique'' (CNRS, also under the PICS action 4263).
It has received funding from the European Union's Horizon 2020 research and innovation program under the Marie Sk{\l}odowska-Curie grant agreement n\textsuperscript{o} 642294.
This work was also supported by the ``Programme Investissements d'Avenir'' under the
program ANR-11-IDEX-0002-02, reference ANR-10-LABX-0037-NEXT.
The calculations of this work have been partly performed by using the resources of the HPC center CALMIP under the grant 2016-p1048.

\bibliographystyle{spphys}       

\newpage
\section*{Figures}

\end{document}